\documentstyle[12pt]{article}
\newcommand{\lsim}{\raisebox{0.3mm}{\em $\, <$} \hspace{-3.3mm}
\raisebox{-1.8mm}{\em $\sim \,$}}
\newcommand{\gsim}{\raisebox{0.3mm}{\em $\, >$} \hspace{-3.3mm}
\raisebox{-1.8mm}{\em $\sim \,$}}
\newcommand{\re}{\par\hangindent=0.5cm\hangafter=1\noindent}
\newcommand{\vsp}{\vspace{18pt}}
\begin{document}

\baselineskip 24pt


\begin{center}
\LARGE \bf
\baselineskip 40pt

UV and X-ray Spectral Lines of FeXXIII Ion 
for Plasma Diagnostics


\normalsize \rm
\vspace{12pt}
Izumi Murakami and Takako Kato   
\vspace{12pt} \\
 National Institute for Fusion Science, Nagoya 464-01, Japan
\vsp

\vspace{2cm}

\vspace{12pt}
{\bf ABSTRACT}
\end{center}

\baselineskip 24pt
\vspace{5pt}

We have calculated X-ray and UV spectra of Be-like Fe (FeXXIII) ion
in collisional radiative model  including
all fine-structure transitions among the  $2s^2$, $2s2p$, $2p^2$,
$2snl$, and $2pnl'$ levels where $n=3$ and 4,
adopting data for
the collision strengths
by Zhang \& Sampson (1992) 
and by Sampson, Goett, \& Clark (1984). 
Some line intensity ratios can be used for the temperature 
diagnostics.
We show 5 ratios in UV region and 9 ratios in X-ray region as a function
of electron temperature and density
at  $0.3 {\rm keV} \lsim T_e \lsim 10 {\rm keV}$ 
and $ n_e = 1 - 10^{25} \rm cm^{-3}$.
The effect of cascade in these line ratios 
and in the level population densities
are discussed.
\\

\vspace{2cm}
\begin{center}
Accepted by {\it Physica Scripta}
\end{center}

\newpage


\section{Introduction}

Intensity ratios of emission lines of ions can be used
for plasma  diagnostics to investigate 
physical condition of plasmas.
FeXXIII, Be-like Fe ion, is observed in fusion plasma
at around $T_e \sim $1keV as well as
astrophysical plasmas, such as the solar corona and flares [1],
and the intracluster gas [2]. 

Historically, so-called corona model has been used to
calculate level population densities to get 
intensities of emission lines, considered 
only collisional excitation from the ground state in the
calculation.
However, the metastable states play an important role
for the population density of L-shell ions
even at low density region.

We have constructed collisional radiative model for L-shell ions,
including all transitions up to $n=4$.
This model takes into account the effects of 
excitation from metastable states
and radiative cascades from the upper levels.

Bhatia \& Mason [3]  examined the population densities 
of FeXXIII, including all transitions
among 20 levels, $2s^2$, $2s2p$, $2p^2$, and $2s3l$, and
estimated line intensity ratios   as a function of electron temperature. 
Keenan et al. [4]  examined the intensity ratios 
for transitions between $n=2$ levels of FeXXIII including 10 levels.
Since the number of levels considered in their calculations
is small, the effect of radiative cascade to the population densities
is underestimated.

Here, we calculate the population density of FeXXIII  ion
by the collisional radiative model, including all transitions
among 98 levels of
$2s^2$, $2s2p$, $2p^2$, $2snl$, and $2pnl$ ($ n \leq 4$)
with different data from the previous authors
for collision strengths 
given by Zhang and Sampson [5], and
Sampson, Goett, and Clark [6].
Using the results, we study the line intensity ratios as a function
of electron temperature and density for plasma diagnostics.
We discuss the effect of metastable state and
radiative cascade from the upper levels
on the population density as well as the line intensity ratio.

\section{Atomic Data}

The fine structure energy levels and transition probabilities
has been calculated by AUTOLSJ method in Dubau et al. [7] 
by SUPERSTRUCTURE Code 
for the principal quantum number $n \le 4$, and
98 levels of the $1s^2 2snl$ and $ 1s^2 2pnl$ 
configurations are considered.
We have also included a magnetic quadrapole transition probability
for $2s^2 \ ^1S_0 - 2s2p \ ^3P_2 $   
           given by Nussbaumer \& Storey [8], 
magnetic dipole transition probabilities 
for $2s2p ( ^3P_0 - \ ^3P_1) $ and
for $2s2p ( ^3P_1 - \ ^3P_2) $ transitions, 
and an electric quadrupole transition probability
for $2s2p ( ^3P_0 - \ ^3P_2) $ transition  
given by Bhatia and Mason [3].

We have adopted electron collision strengths calculated
by Zhang \& Sampson [5]  
for the  $\Delta n=0$ transitions with n=2,
and those by Sampson, Goett, \& Clark [6] 
for all fine-structure transitions between $n=2$ and $n=3$,
to get the excitation rate coefficients.
We have used modified Mewe's empirical formula [9] 
for other transitions.
For the parameter A of Mewe's Gaunt factor
we have adopted $A=0.075$ which is
half of original value
for allowed transitions.
The excitation rate coefficients calculated by Mewe's
empirical formula 
are checked for available transitions with above data.
We have included the proton collisional excitation
for the three fine-structure transitions in the
$2s 2p \ ^3P$ levels, using the cross sections given by Doyle [10].

Zhang \& Sampson  calculated the collision strengths  
with  relativistic distorted-wave program.
Using their data provided as a table,
we have calculated the effective collision strength, $ \gamma$,
by approximating the integration with
the Maxwellian distribution of electron velocity to a summation
used the orthogonal Laguerre polynomials
(Formula 25.4.45 in Handbook of Mathematical Functions [11]).
Sampson et al.  used  
a relativistic Coulomb-Born-exchange method
and gave collision strengths fitted by an analytical form.

The collision strengths used by other authors [3,4] are calculated with
different methods.
Bhatia \& Mason [3] calculated the collision strengths using distorted wave
approximation for the transitions between 20 levels.
Norrington \& Grant [12] used a relativistic R-matrix code
to calculate the collision strengths for transitions
among 10 levels, which were used by Keenan et al.[4].

\section{Collisional Radiative Model}

We have calculated 
the level population densities and spectrum
by Collisional-Radiative Model (CRM). 
We have taken 
into account the collisional excitation between all
fine-structure transitions including forbidden transitions
as well as the radiative transitions;
\begin{equation}
\frac{d n_i}{dt} =
 - \sum_j ( C_{ij} n_e + A_{ij} ) n_i
 + \sum_j  C_{ji} n_e  n_i
 + \sum_{j>i}  A_{ji}  n_i
 = 0,
\end{equation}
where $C_{ij}$ is the excitation rate coefficients from the level i to j,
$A_{ij}$ is the radiative transition probability from i to j,
$n_i$ is the population density of i level and
$n_e$ is the electron density.
We assume the equilibrium, $dn_i/dt=0$,
 to get the population density,
given the electron density and temperature.
Here we do not take into account ionization and recombination
processes from and/or to other Fe ions.
The line intensity per electron and per ion
from the level i to j  is given as
\begin{equation}
I_{ij}/n_I n_e = A_{ij} (n_i/ \sum_k n_k) / n_e,
\end{equation}
where $n_I \equiv \sum n_k$ is to normalize the population density,
and $n_e$ is the electron density.
We call $I_{ij}/n_In_e$ 
the effective emission rate coefficient.

\subsection{Population Densities}

Be-like ions have metastable state of $2s2p \ ^3 P_J$ from which 
the transition probabilities to
the ground state, $2s^2 \ ^1S_0$, are quite small.
Since the probability, $A_r(2s2p \ ^3P_0 - 2s^2 \ ^1S_0)$ is zero,
the population of the $2s2p \ ^3P_0$ level is not zero
even at low density.
The reduced population density, $n(i)/g(i)/n(1)$
where $g(i)$ is the statistical weight of the level i,
of $2s2p \ ^3P_0$ level is constant, 0.007 for $T_e$ = 1 keV,   
at $n_e \lsim 10^{12} \rm cm^{-3}$
and increase above $10^{12} \rm cm^{-3}$.
At $10^{14} \lsim n_e \lsim 10^{16} \rm cm^{-3}$
the reduced population density of
other metastable state, $2s2p \ ^3P_2$, approaches the
population density of $2s2p \ ^3P_0$ level and both 
are nearly constant.
And above $10^{17} \rm cm^{-3}$ all three metastable levels,
$2s2p \ ^3P_{J=0,1,2}$, have almost the same reduced
population density.
For other levels the population densities increase
proportionally to the electron density at $n_e \lsim 10^{12} \rm cm^{-3}$.
Above $10^{12} \rm cm^{-3}$, $2p^2$ levels and some triplet levels of
$n=3$ and 4 have the population densities which vary according to the
variation of the population densities of the metastable states.
The reduced population densities of all $n=2$ levels become constant
at $n_e \sim 10^{21} \rm cm^{-3}$
and those of all $n=3$ and $n=4$ levels become constant
at $n_e \sim 10^{23} \rm cm^{-3}$,
where these levels
are to be populated proportionally to the
statistical weights 
[14].

\subsection{Emission Line Intensities}

Figure 1 shows
the effective emission rate coefficients 
defined by equation (2)
for  17 selected transitions
as a function of electron temperature
at electron density of $10^{11} \rm cm^{-3}$,
where there is no density effect on line intensities.
The transitions are listed in Table 1.

The rate coefficients
of transitions among $n=2$  levels in Figure 1a
are monotonic functions of electron temperature, $T_e$.
We find how the cascades affect the line intensities at
$ T_e \gsim 10^7$K.
In Figure 1a, thin lines represent the line intensities calculated
with 10 levels, i.e. only $n=2$ levels (case A) has been made by Keenan et al. [4].
Thick lines are calculated with 46 levels, up to $n=3$ levels (case B),
and the thickest lines are calculated with 96 levels, up to
$n=4$ levels (case C).
Each line is labeled with number in Table 1 and A, B, or C indicating
case A, B, or C, respectively.
The line intensities of case A are almost dominated by the
collisional excitation from the ground state
and monotonically decrease as electron temperature increases.
The differences between thin lines (case A) and thick lines 
(case B and C) are prominent at $T_e \gsim 10^7$K, which
caused by the radiative cascades from the upper levels.
In particular, $2p^2$ levels are populated by the cascades considerably and
the line intensities increase as the electron temperature increases.
For instance,  in case A the line intensity of $ 2s2p \ ^1P - 2p^2 \ ^1S$
transition ($L_4$) is only 10 \% of that in case C 
at $T_e=10^8$K ($\sim $ 10keV).
The difference between cases B and C is not so large ($1 \sim 14$ \%) even 
at $T_e=10^8$K.
Line intensities at $n_e=10^{14} \rm cm^{-3}$ in case C are also plotted
with dash-dashed lines labeled as 1' in Figure 1a.
$L_1$, $L_2$, and $L_3$ are smaller than those at 
$n_e=10^{11} \rm cm^{-3}$ since the population densities
divided by $n_e$ for the metastable states are smaller
due to collisional excitation from these levels to upper levels.
The intensities, $L_5$ and $L_6$, become larger
because of the same reason. 

Figure 1b shows
the effective emission rate coefficients
of $n=3-2$ transitions 
as a function of electron temperature, $T_e$.
In the temperature range of $3 \times 10^6 - 10^8$K
(0.3keV - 10keV), some
intensities increase as temperature increases and some
have a maximum.
Such different temperature dependences allow us
to use line intensity ratios as an indicator
of electron temperature.
Similarly to $n=2-2$ transitions, the effect of the radiative cascade
is seen in the figure.
The differences of case B and C become larger at high temperature
and are 
 1\% -- 28\% at $1.2 \times 10^8$K (10keV), depending on transitions.
$2s3d$ levels are much affected by the cascade.

Figure 1c shows the effective emissions rate coefficients
of $n=4-2$ transitions.
The temperature dependences are similar to $n=3-2$ transitions.
Note that in the calculation of population densities
the excitation rate coefficients 
from $n=2$ to $n=4$ transitions are estimated 
 by modified Mewe's formula mentioned in Sec.2.

\section{Line Intensity Ratios}

We have chosen 17 strong lines to measure the electron temperature of a plasma,
which are summarized in Table 1.
Five ratios from six
lines chosen among $n=2-2$ transitions 
 in UV spectrum, are taken to be examined,
as selected by Keenan et al. [4]. 
Six ratios from seven lines
chosen among $n=3-2$ transitions at
around 11 \AA \ in X-ray spectrum, and
three ratios from four lines
chosen among $n=4-2$ transitions at around 8 \AA \ are considered.
We summarize the line ratios in Table 2.

\subsection{Density Dependences}

First, we investigate the density dependence of the above line ratios.
Figure 2 shows the line ratios listed in Table 2
as a function of electron density, $n_e$,
at electron temperature of $1.2 \times 10^7$K (1keV).
There is no density dependence seen at $n_e \lsim 10^{12} \rm cm^{-3}$.
$R_6$ and $R_{12}$ are relatively insensitive to the
density in the whole range, while
$R_2$, $R_3$, $R_5$, $R_9$, and $R_{10}$
 vary above $n_e \gsim 10^{12} \rm cm^{-3}$.
Their density dependences are caused by the variation of population densities
of metastable levels.
The variation of $R_5$
at $10^{12} \lsim n_e \lsim 10^{16} \rm cm^{-3}$ is caused by the
population density of $2s2p \ ^3P_2$ level.
The population density becomes nearly constant
and the effective emission rate coefficient of the line, $L_3$,
decreases proportionally to the electron density (see eq.(2)).
Meanwhile the population density of $2s3p \ ^3P_1$ level 
increases proportionally
to the electron density and the effective emission rate coefficient of
$L_2$ keeps constant. 
At this density region, the variation of 
the population densities of $2s2p \ ^3P_{J=0,2}$ levels
affects the population densities of $2p^2 \ ^3P_2$ and $ ^1D$ levels,
and then the line ratios, $R_2$ and $R_3$, vary.
The variation of $R_1$ - $R_4$ 
at $10^{17} \lsim n_e \lsim 10^{21} \rm cm^{-3}$ is caused by the
decrease of the effective emission rate coefficient of $L_2$ line,
because the population density of $2s2p \ ^3P_1$ level increases
slower than the increase of the electron density.
Other ratios are almost constant at $n_e \lsim 10^{16} \rm cm^{-3}$.

For tokamak plasmas at electron density around $10^{13}-10^{14} \rm cm^{-3}$
the line ratios, 
$R_2$, $R_4$, $R_5$, $R_9$, and $R_{10}$,
are not good for temperature diagnostic
because of strong density dependences.
Checked the temperature dependences of these ratios in next section,
the ratio $R_5$ is found to be quite good for density diagnostic
because of the small temperature dependence.
Other ratios have possibilities to be used for temperature diagnostic.

\subsection{Temperature Dependences}

Figure 3 shows the line intensity ratios
as a function of electron temperature at $n_e=10^{11} \rm cm^{-3}$. 
For $R_1$ to $R_5$, we plot calculated 
ratios, including 10 levels (case A), 
with 46 levels (case B), 
and with 98 levels (case C), 
together with the ratios obtained by Keenan et al.  [4].
For $R_6$ to $R_{10}$,
ratios in cases C and B are plotted.
Crosses  in Figures 3(e)-(h)
are ratios estimated from results by Bhatia and Mason [3]
at an electron density of $10^9 \rm cm^{-3}$.
There is no density dependence for these line ratios
at this density range.
Bhatia and Mason obtained these ratios with 20 levels. 
When we calculated the ratios with the same 20 levels,
such line ratios are different by 1 \% or less from the ratios
calculated in case B (46 levels)
in which $2pnl$ levels are included. 
So we can compare their ratios with ours in case B.
Those ratios at $n_e=10^{14} \rm cm^{-3}$ in case C are also plotted
in the figure.

Comparing $R_1 - R_5$ in case A with the ratios obtained
by Keenan et al., we find that the differences are 1 \% - 33 \%
and tend to increase as temperature increases.
The ratios, $R_3$ and $R_4$, are much different from Keenan et al.'s results. 
The difference is
caused by the different excitation cross sections
from the ground state to $2p^2$ states.

When we include $n=3$ and 4 levels in the calculations, the line ratios
are significantly changed, especially at higher temperatures.
It is because of the influence of the cascades from the upper levels contributing to the
population densities and line intensities, as seen from Figure 1a.
For example, $R_2$ in case C is 93 \% of $R_2$ in case A 
at $T_e= 3.5 \times 10^6$K (0.3keV),
57 \% at $T_e= 1.2 \times 10^7$K (1keV), 
and 32 \% at $T_e=1.2 \times 10^8$K (10keV).
Similarly to the line intensities themselves, the intensity ratios
are much affected by the cascade.
Because 
the population densities of the metastable states depend on
the density at $n_e \lsim 10^{12} \rm cm^{-3}$ and
the collisional excitation from the metastable states starts to affect
the population densities of upper levels,
 $R_2$, $R_3$, and $R_5$ show
different temperature dependences at $n_e=10^{14} \rm cm^{-3}$.

As seen in Figures 3(e)-(h), the ratios, $R_6$-$R_{10}$, are mostly
different from those obtained by Bhatia and Mason.
$R_7$ shows good agreement with that of Bhatia and Mason 
within 10 \% difference.
In particular,
$R_6$ and $R_8$ in case B are different by 17-48 \% from Bhatia and Mason.
Those differences are caused by difference of the excitation cross section.
The collision strength of $2s^2 \ ^1S-2s3d \ ^1D$ obtained by
Sampson et al. increases monotonically as the incident electron energy
increases, while that of Bhatia and Mason decreases at higher
energy.
In addition the collision strength of Sampson et al. is always larger
than that of Bhatia and Mason.
These two effects increase the population density of $2s3d \ ^1D$,
resulting in different value for $R_6$
between Bhatia and Mason and ours.
For $R_8$, the collision strengths of
$2s^2 \ ^1S - 2s3p \ ^1P$ and $2s^2 \ ^1S - 2s3p \ ^3P_1$
obtained by Sampson et al. increase similarly 
according to the increase of the incident electron
energy, while the collision strength of $2s^2 \ ^1S - 2s3p \ ^1P$ 
obtained by Bhatia and Mason 
increases faster than that of $2s^2 \ ^1S - 2s3p \ ^3P$. 
This causes the different temperature dependences for $R_8$.
The ratios of
$R_9$ and $R_{10}$ decrease slower than those of
Bhatia and Mason at higher temperature.

The effect on the line ratios by including $n=4$ levels is
small, by amount up to  25 \%,
for all ratios considered here.
We may note that it is important to include levels
up to $n=3$ at least
in calculations with CRM 
to get reliable line intensity ratios
for temperature diagnostics.

As $R_8$ and $R_{11}$ show poor temperature dependence, 
a combined usage of these
ratios with others is quite helpful to 
check the calibration of spectrometer.
$R_9$ and $R_{10}$ depend slightly on electron density.
$R_9$ at $10^{14} \rm cm^{-3}$ is 1.6-2.4 times larger than  that
at $10^{11} \rm cm^{-3}$, and $R_{10}$  at $10^{14} \rm cm^{-3}$
is 1.2 times larger that at $10^{11} \rm cm^{-3}$.

Although the line intensities of $n=4-2$ transitions are weak,
the wavelength region is not so crowded and these lines
is expected to be useful for plasma diagnostics
because of being unblended.
$R_{12}$ is sensitive to electron temperature similarly to $R_{11}$,
while $R_{13}$ is quite insensitive to temperature.

\section{Discussion}

\subsection{Contribution of Cascade and Metastable State}

We have found that the cascades are important for dominating 
an emission line intensity. 
The effect of the cascade can be examined with population 
densities which are analytically calculated with following equation (3) .
Only collisional excitation from the ground state and
radiative transitions are considered in the equation.
Neglected transitions are unimportant at $n_e \lsim 10^{15} \rm cm^{-3}$.
\begin{equation}   \label{eq-cascade}
\frac{n_i}{n_1} = \frac{ C_{1i} n_e}{\sum_{j=1}^{i-1} A_{ij} }
 + \sum_{k=i+1}^{m} \frac{n_k}{n_1} \frac{ A_{ki}}{\sum_{j=1}^{i-1} A_{ij} },
\end{equation}
where level 1 means the lowest level and level m means the highest level,
which is 98 in this case.
The first term in the right hand side  
is the contribution of the collisional excitation 
from the ground state
and the second term represents the contribution of the cascades.
This equation can be calculated from the highest level for descending $k$.
Similarly, we can also examine the effect of the metastable state 
to the population densities separately,
changing the lowest level in equation (3) to be the metastable state,
$2s2p \ ^3P_0$.

Comparing the population densities obtained with equation (3)
and those by CRM with equation (1), 
we find that the cascade from $n=3$ to $n=2$ levels are important
to the population densities of $n=2$ levels, especially $2p^2$ state.
For example, Figure 4 shows each contribution of 
the collisional excitation from the ground state (white in the figure),
of the cascade (vertical-striped, from n=4 levels;
dotted, from n=3 levels;
lateral-striped, from n=2 levels), and of
the metastable state (black),
to population density for each level
in the plasma at $n_e=10^{11} \rm cm^{-3}$ and 
$T_e=1.2 \times 10^7$K (1keV).
Sixty \% of the population density of $2p^2 \ ^1S_0$ level 
and
17 \% of that of $2s2p \ ^3P_1$ level are fed by the cascade.
Cascade from $n=4$ to $n=2$  levels contributes the
population densities by  1-10 \% to $n=2$ levels
and is also effective for $2p3l$ levels and 
$2s3l$ triplet levels. 
Such percentages depend on the electron density and temperature.
The results are consistent with the arguments on the
differences of line intensities in cases A, B, or C, which
include different number of levels in the model
(\S 3.2).

The contribution originated from the metastable state 
to the population densities
is mostly small, 
 however, some triplets and $2p3l$ levels,
such as $2p^2 \ ^3P_1$, $2p3s \ ^3P_0$, $2p3p \ ^1P$,
 and $2p3d \ \ ^3D_2$,  have 
more than 50 \% contribution
from the metastable state.
The contributions of the metastable state
to $L_2$, $L_3$, $L_5$, $L_6$, $L_{11}$, and $L_{12}$
line intensities are 6\%, 10\%, 6\%, 4\%, 2\%, and 2\%, respectively,
at $n_e=10^{11} \rm cm^{-3}$ and $T_e=1.2 \times 10^7$K (1keV).
The contributions to other lines are much less than 1\%.
At higher densities,  the collisional excitation from excited states
cannot be neglected 
and the equation (3) should not be used.

\subsection{Comparison with Observations of the Solar Flares}

There are some observations of FeXXIII lines in the Sun.
Mason et al. (1984) [14] observed a solar flare with {\it OSO 5}
satellite and measured the intensity ratio, $R_1$, from the spectrum.
They obtained $R_1 \simeq 23$ with $\sim 25$ \% uncertainty.
Keenan et al. discussed that their theoretical result was in good
agreement with the observation of Mason et al.
Our ratio, $R_1$, in case C is  different by 4-10 \% from 
the ratio of Keenan et al., however, it still agrees with the
observation within the uncertainty, at the temperature where
Be-like Fe (FeXXIII) ion has the maximum fractional abundance in ionization
equilibrium ($T_e=10^{7.1}$K ($\sim$ 1keV); Arnaud \& Rothenflug 1985 [15]).
Our ratio gives a larger temperature than the ratio of Keenan et al. did.
Because of the measurement error it is not possible to determine the
electron temperature of the solar flare in this case.

McKenzie et al. (1985) [1] observed a solar flare
with {\it SOLEX} and showed the emission lines in the 5.5 - 12 \AA \
range.
They detected 4 lines of $n=3-2$ transitions and 2 lines of $n=4-2$
transitions.
Since the lines at $\sim$ 11 \AA \ are blended and the background level
is not easy to determine, their line intensities
have a large uncertainty.
The ratio, $R_6$, is measured as 2.04 by McKenzie et al.
The authors concluded that the ratio was in good agreement
with one estimated by Bhatia and Mason.
In our case, this ratio indicates $T_e \sim 10^{7.6 \pm 0.3}$K,
which is higher than the temperature, $10^{7.1 \pm 0.2}$K, given
by the ratio of Bhatia and Mason.
On the other hand, the ratio, $R_{12}$, from $n=4-2$
transitions is measured as 0.89 by McKenzie et al., and
this indicates $T_e \sim 10^{7.1}$K.

In future {\it SOHO} satellite will observe solar FeXXIII
lines and will provide us more observational data to
be compared with theoretical calculations.

\section{Concluding Remarks}

We study the spectrum of Be-like Fe
(FeXXIII) ion for plasma diagnostic
and show some useful pairs of the emission lines
for temperature diagnostic.
In the collisional-radiative model we have included all fine structure
levels up to $n=4$ (98 levels).
This is new work because previous work for plasma diagnostics
includes levels of up to $2s3l$ (20 levels) [3].
We find the importance of the cascade from upper levels
to the population density. 
However, in this study we do not take into account the
effects of ionization and recombination to/from other ions.

Radiative and 
dielectronic recombination from Li-like Fe (FeXXIV) ion will contribute
much to population densities as like radiative cascade,
especially at low temperature.
For example, total dielectronic recombination rate coefficient is
about $ 7 \times 10^{-12} \rm cm^{3}s^{-1}$ at $T_e=1.2 \times 10^7$K (1keV)
(Moribayashi et al. 1995 [16])
and the process will affect weak lines such as
$L_4$-$L_6$, $L_{10}$-$L_{17}$. 
The recombination will increase 
the line intensity of $L_2$,
$2s^2 \ ^1S-2s2p \ ^3P_1$, by 13 \%   
at the maximum
when ion abundances of Li-like (FeXXIV) and Be-like (FeXXIII) ions
  are assumed to be equal.

Inner-sub shell ionization, $ 2s^2 2p + e \rightarrow 2s2p + 2e$,
 from B-like Fe (FeXXII) will make
the population density of metastable state increase,
which will affect the densities of other levels
(Kato et al. 1995 [17]).
The inner-sub shell
ionization rate coefficient  can be estimated as
$6.3 \times 10^{-12} \rm cm^3s^{-1}$  at $1.2 \times 10^7$ K (1keV)
with Lotz's formula [18].
If the rate coefficient to $2s2p \ ^3P_0$ level is estimated with
the above value
weighted with the statistical weight,
the population density of the level would increase by 25\% with
this process, compared with the excitation rate coefficient from
the ground state.
Here we assume
that ion abundances of B-like (FeXXII) and Be-like (FeXXIII) ions  are equal.
The contribution of the line intensity of $L_2$ will be 3\%.
At higher temperature or with the suprathermal electrons
the contribution will increase and the
process will be more important.
Therefore
we need to include such effects to the calculation in  the future work.
\\

\vspace{12pt} \noindent
{\bf Acknowledgments}

We acknowledge H.L.Zhang for providing us his data
of the collision strengths for n=2 - n = 2 transitions.

\vspace{24pt}
\noindent
{\large \bf References}

\re
[1] McKenzie, D.L., Landecker, P.B., Feldman, U., and Doschek, G.A.,
Astrophys..J., {\bf 289},  849 (1985)

\re
[2] Fabian, A.C., Arnaud, K.A., Bautz, M.W., and Tawara, Y.,
Astrophys.J., {\bf 436},  L63 (1994)

\re
[3] Bhatia, A.K. and Mason, H.E., Astron. Astrophys., {\bf 103},  324 (1981)

\re
[4] Keenan, F.P., Conlon, E.S., Warren, G.A., Boone, A.W., and Norrington, P.H.,
Astrophys.J., {\bf 405},  350 (1993)

\re
[5] Zhang, H.L. and Sampson, D.H.,
Atomic Data and Nuclear Data Tables, {\bf 52},  143 (1992)

\re
[6] Sampson, D.H., Goett, S.J., and Clark, R.E.H.,  
Atomic Data and Nuclear Data Tables, {\bf 30},  125, (1984)

\re
[7] Dubau, J., Cornille, M., Bely-Dubau, F., Faucher, P., and Kato, T., 
``New Horizon of X-ray Astronomy - First Results from ASCA,
Frontiers Science Series No. 12 " 
(Universal Academy Press, Tokyo), 615 (1994) 

\re
[8] Nussbaumer, H. and Storey, P.J.,
J.~Phys.~B., {\bf 12},  1647, (1979)

\re
[9] Mewe, R., Astron. Astrophys., {\bf 20},   215 (1972)

\re
[10] Doyle, J.G.,
Atomic Data and Nuclear Data Tables, {\bf  37},  441 (1987)

\re
[11] {\it HandBook of Mathematical Functions} 
ed. M.~Abramowitz and I.~A.~Stegun
(New York, Dover), p.890 (1965).

\re
[12] Norrington, P.H. and Grant, I.P., J.~Phys.~B, {\bf 20},  4869 (1987)

\re
[13] Murakami, I., Kato, T., and Dubau, J., Research Report of the  
National Institute for Fusion Science, NIFS-DATA-35 (1996)

\re
[14] Mason, H.E., Bhatia, A.K., Kastner, S.O., Neupert, W.M., \&
Swartz, M., Sol. Phys., {\bf 92},  199 (1984)

\re
[15] Arnaud, M. \& Rothenflug, R., 
Astron. Astrophys. Suppl. {\bf 60},  425 (1985)

\re
[16] Moribayashi, K., Kato, T., \& Safronova, U.,
submitted to Fusion Engineering and Design
(1995)

\re
[17] Kato, T. et al.
submitted to Fusion Engineering and Design
(1995)

\re
[18] Lotz, W., Astrophys.J.Suppl., {\bf 14},  207 (1967)

\newpage
\noindent
{\bf Figure Captions}
\\

\noindent
{\bf Fig. 1}.
Effective emission rate coefficients 
as a function of electron
temperature at an electron density of $10^{11} \rm cm^{-3}$.
Horizontal axis is for the electron temperature measure in K (lower axis)
or keV (upper axis).
(a) Coefficients for transitions 
among $n=2$  levels.
The heaviest lines are calculated in case C including 98 levels up to $n=4$,
heavy lines are
in case B including 46 levels up to $n=3$, and thinest lines are in case A
including 10 levels of $n=2$.
Each line is labeled with numbers in Table 1 and A, B, or C for cases
A, B, or C, respectively. 1A means $L_1$ lines in case A, and so on.
Long dash-short dashed lines labeled with number with prime such as 1'
are in case C at an electron density of $10^{14} \rm cm^{-3}$.
The line 4' is overlapped with the line 4C
at $n=10^{11} \rm cm^{-3}$;
(b)
For transitions from $n=3$ to $n=2$ levels.
Heavy lines are calculated in case C and thin lines are in case B. 
Similar to Fig.1a, each line is labeled with number in Table 1 ; and 
(c)
For transitions from $n=4$ to $n=2$ levels.
Similar to Fig.1a, each line is labeled with number in Table 1.
\\

\noindent
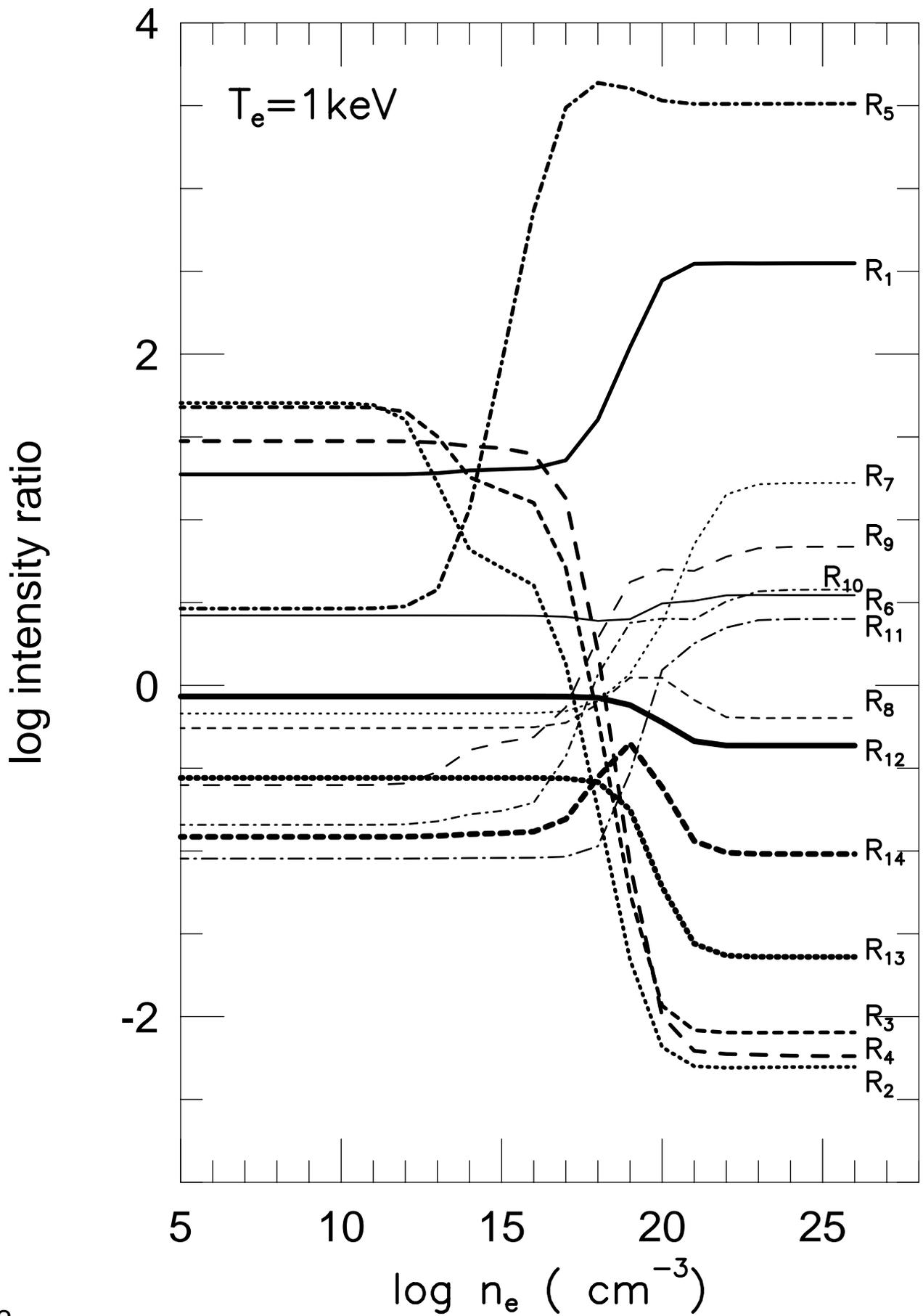
{\bf Fig. 2}. 
The theoretical FeXXIII  emission-line ratios as a function
of electron density at electron temperature of
$T_e=1$ keV.  
Fourteen ratios from $R_1$ to $R_{14}$  listed in Table 2 are plotted.
\\

\noindent
{\bf Fig. 3}. 
The theoretical FeXXIII  emission-line ratios as a function
of electron temperature at an electron density of
$n_e=10^{11} \rm cm^{-3}$.
Ratios are listed in Table 2.
The label A indicates ratio calculated in case A including 10 levels of $n=2$,
 B indicates case B including 46 levels up to $n=3$,
and C indicates case C including 98 levels up to $n=4$. 
Ratios at $n_e=10^{14} \rm cm^{-3}$ in case C are also plotted
with dotted lines, some of which are overlapped with lines
in case C at $n=10^{11} \rm cm^{-3}$.
Lines labeled with K in panels (a)-(d) are results by Keenan et al.[4] and
crosses in panels (e)-(h) are results by Bhatia and Mason [3].
\\

\noindent
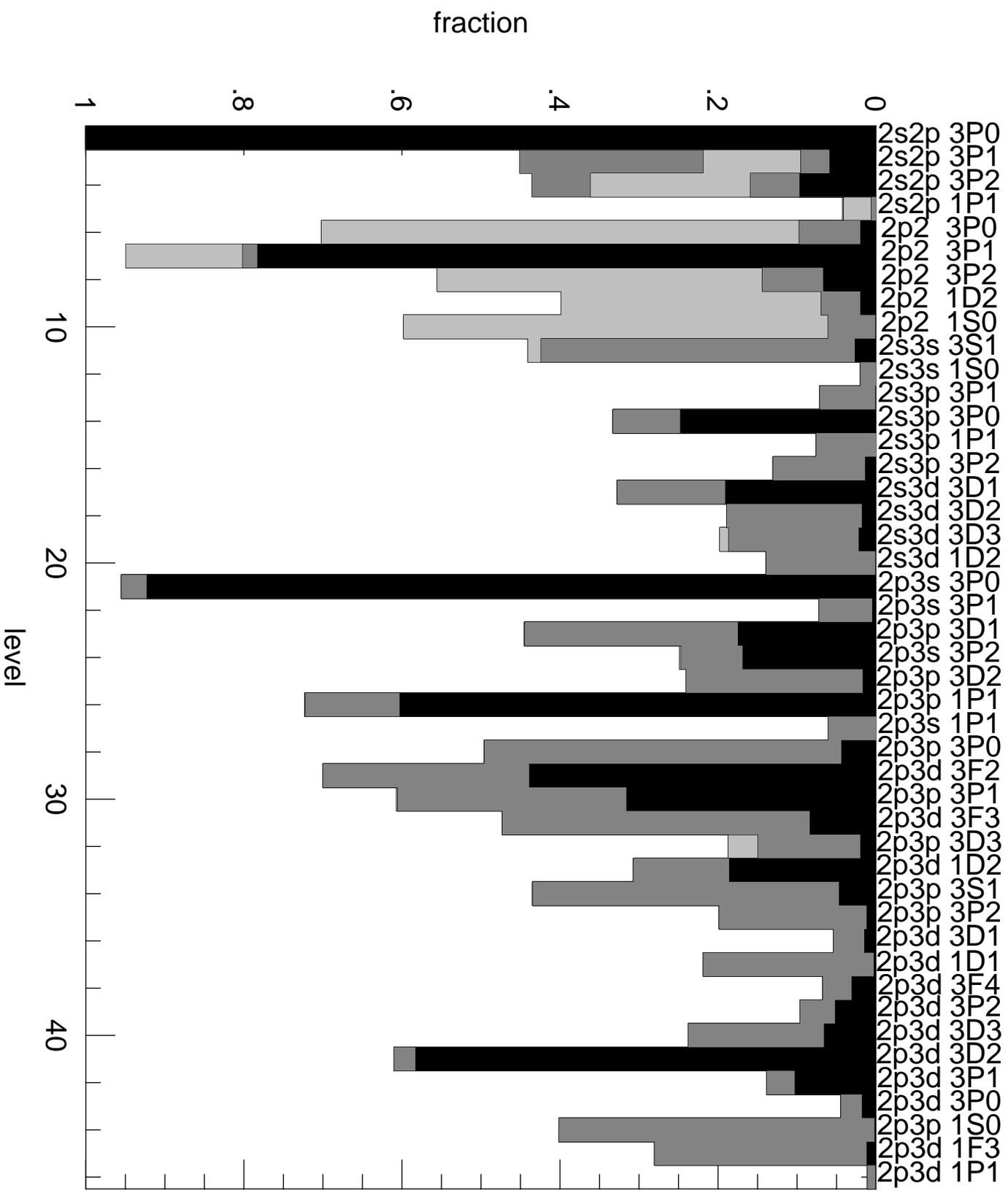
{\bf Fig. 4}.
The fraction of each contribution to population density of
each level at $n \leq 3$, at $T_e=1$keV and $n_e=10^{11} \rm cm^{-3}$.
White shows the contribution of the collisional excitation
from the ground state. 
The contributions of
cascade from $n=2$ levels, $n=3$ levels, and $n=4$ levels,
are shown by
lateral-striped, dotted, and vertical-striped, respectively.
Black shows the contribution of both
collisional excitation and cascade originated from the metastable state,
$2s2p \ ^3P_0$.

\newpage

%
%
\begin{table}[t]                          
\caption{Strong Emission Lines of FeXXIII Ions}
\begin{center}
\begin{tabular}{lllrl}
\hline \hline 
label &  &  transitions & wavelength (in \AA) & (in eV) [7]\\  \hline
$L_1$ & $\cdots$ & $2s^2 \ ^1S - 2s2p \ ^1P$ & 131.48 \AA  & 94.296eV  \\
$L_2$ & $\cdots$ & $ 2s^2 \ ^1S - 2s2p \ ^3P_1$ & 267.62 \AA  & 46.328eV  \\
$L_3$ & $\cdots$ & $2s2p \ ^3P_1-2s2p \ ^3P_2 $ & 976.78 \AA &12.693eV \\
$L_4$ & $\cdots$ & $2s2p \ ^1P -2p^2 \ ^1S $ & 146.21 \AA & 84.800eV  \\
$L_5$ & $\cdots$ & $2s2p \ ^1P -2p^2 \ ^1D $ & 219.43 \AA & 56.502eV  \\
$L_6$ & $\cdots$ & $2s2p \ ^3P_2-2p^2 \ ^3P_2$ & 167.33  \AA & 74.097eV  \\
  &  &  &  &  \\
$L_7$ & $\cdots$ & $2s2p \ ^1P-2s3d \ ^1D$ & 11.725 \AA & 1057.4 eV \\
$L_8$ & $\cdots$ & $2s2p \ ^1P - 2s3s \ ^1S $ &
        12.149 \AA &  1020.5 eV \\
$L_9$ & $\cdots$ & $2s^2 \ ^1S - 2s3p \ ^1P $ &
                                 10.966 \AA & 1130.7 eV \\
$L_{10}$ & $\cdots$ & $ 2s^2 \ ^1S - 2s3p \ ^3P_1 $ &
   11.006 \AA & 1126.5 eV \\
$L_{11}$ & $\cdots$ & $ 2s2p \ ^3P_2 - 2s3d \ ^3D_3 $ &
       11.429 \AA & 1084.8 eV \\
$L_{12}$ & $\cdots$ & $ 2s2p \ ^3P_1 - 2s3d \ ^3D_2 $ &
       11.308 \AA & 1130.8 eV \\
$L_{13}$& $\cdots$ & $ 2p^2 \ ^1S - 2p3d \ ^1P $ &
   11.869 \AA & 1044.6 eV  \\
  &  &  &  &  \\
$L_{14}$& $\cdots$ & $ 2s2p \ ^1P - 2s4d \ ^1D $ &
   8.8074 \AA & 1407.7 eV  \\
$L_{15}$ & $\cdots$ & $ 2s^2 \ ^1S - 2s4p \ ^1P $ &
   8.2934 \AA & 1495.0 eV \\
$L_{16}$& $\cdots$ & $ 2s2p \ ^1P - 2s4s \ ^1S $ &
   8.8998 \AA & 1393.1 eV  \\
$L_{17}$ & $\cdots$ & $ 2s^2 \ ^1S - 2s4p \ ^3P_1 $ &
   8.3059 \AA & 1492.7 eV \\
\hline 
\end{tabular}
\end{center}
\end{table}

%
\begin{table}[t]                          
\caption{Emission Lines Ratios for Plasma Diagnostics}
\begin{center}
\begin{tabular}{llll}
\hline \hline  \\
label & & ratios & with labels in Table 1 \\ \hline

$R_1$ & $\cdots$ & $I(2s^2 \ ^1S - 2s2p \ ^1P)/I(2s^2 \ ^1S-2s2p \ ^3P_1)$
  &  $L_1/L_2$ \\
$R_2$ & $\cdots$ & $I( 2s^2 \ ^1S - 2s2p \ ^3P_1)/I(2s2p \ ^3P_2-2p^2 \ ^3P_2)$
   & $L_2/L_6$ \\
$R_3$ & $\cdots$ & $I( 2s^2 \ ^1S - 2s2p \ ^3P_1)/I(2s2p \ ^1P -2p^2 \ ^1D)$
      & $L_2/L_5$ \\
$R_4$ & $\cdots$ & $I( 2s^2 \ ^1S - 2s2p \ ^3P_1)/I(2s2p \ ^1P -2p^2 \ ^1S)$
       & $L_2/L_4$ \\
$R_5$ & $\cdots$ & $I(2s^2 \ ^1S-2s2p \ ^3P_1)/I(2s2p \ ^3P_1-2s2p \ ^3P_2)$
       & $L_2/L_3$ \\
  &  &  & \\
$R_6$ & $\cdots$ & $I(2s2p \ ^1P-2s3d \ ^1D)/I( 2s^2 \ ^1S - 2s3p \ ^1P)$
    & $L_7/L_9$ \\
$R_7$ & $\cdots$ & $I( 2s^2 \ ^1S - 2s3p \ ^1P)/I( 2s2p \ ^1P - 2s3s \ ^1S)$
     & $L_9/L_8$ \\
$R_8$ & $\cdots$ & $I(2s^2\ ^1S-2s3p\ ^3P_1)/ I( 2s^2 \ ^1S - 2s3p \ ^1P)$
     & $L_{10}/L_9$ \\
$R_9$ & $\cdots$& $I(2s2p \ ^3P_2-2s3d \ ^3D_3)/I( 2s^2 \ ^1S - 2s3p \ ^1P)$
     & $L_{11}/L_9$ \\
$R_{10}$ & $\cdots$& $I(2s2p \ ^3P_1-2s3d \ ^3D_2)/I( 2s^2 \ ^1S - 2s3p \ ^1P)$
     & $L_{12}/L_9$ \\
$R_{11}$& $\cdots$& $I(2p^2\ ^1S-2p3d \ ^1P)/I( 2s^2 \ ^1S - 2s3p \ ^1P)$
     & $L_{13}/L_9$  \\
  &  &  & \\
$R_{12}$ & $\cdots$ & $I( 2s^2 \ ^1S - 2s4p \ ^1P)/I(2s2p \ ^1P-2s4d \ ^1D)$
    & $L_{14}/L_{13}$ \\
$R_{13}$ & $\cdots$ & $I( 2s2p \ ^1P - 2s4s \ ^1S)/I(2s2p \ ^1P-2s4d \ ^1D)$
     & $L_{15}/L_{13}$ \\
$R_{14}$ & $\cdots$ & $I( 2s^2 \ ^1S - 2s4p \ ^3P_1)/I(2s2p \ ^1P-2s4d \ ^1D)$
    & $L_{16}/L_{13}$ \\
\hline 
\end{tabular}
\end{center}
\end{table}

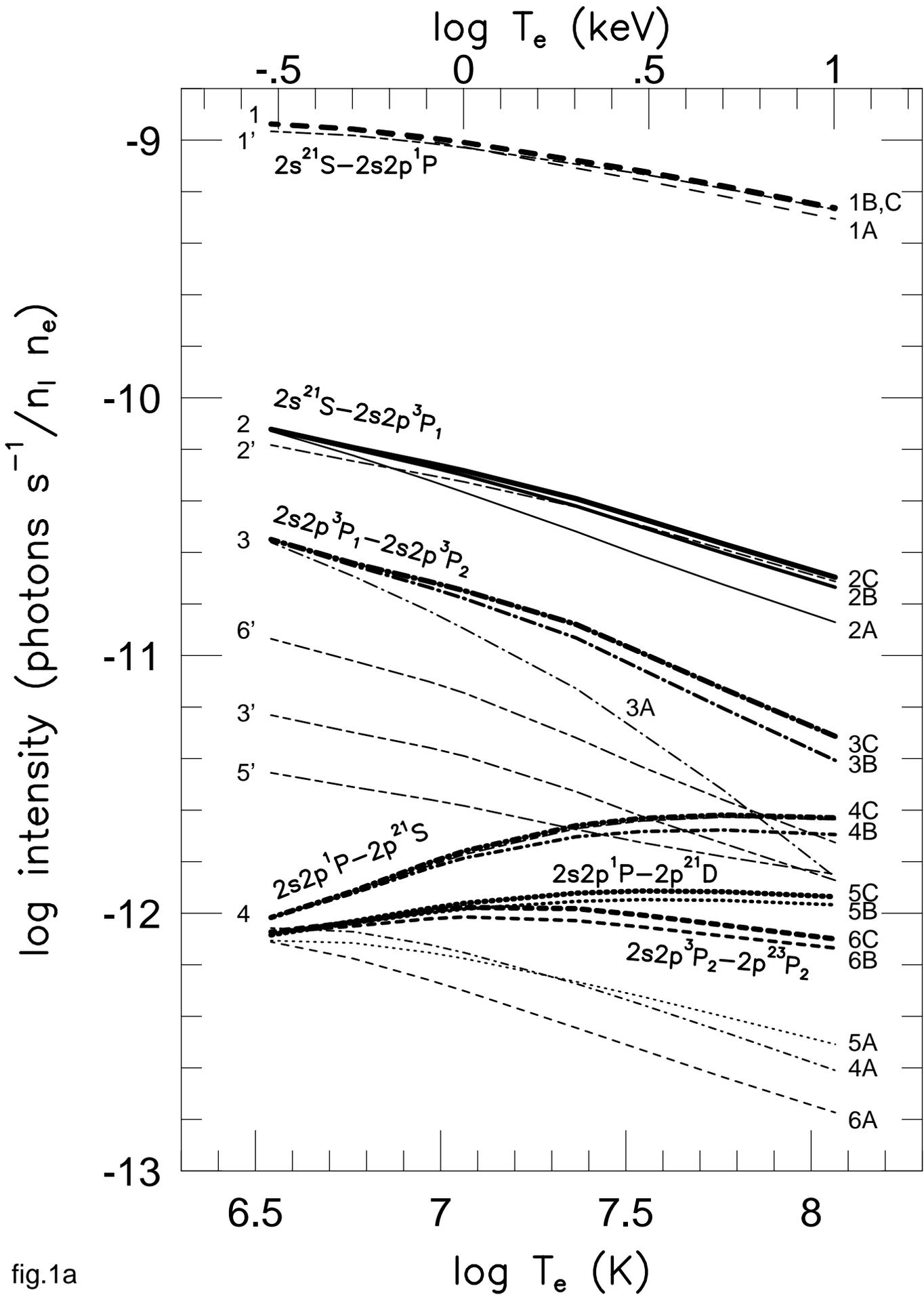
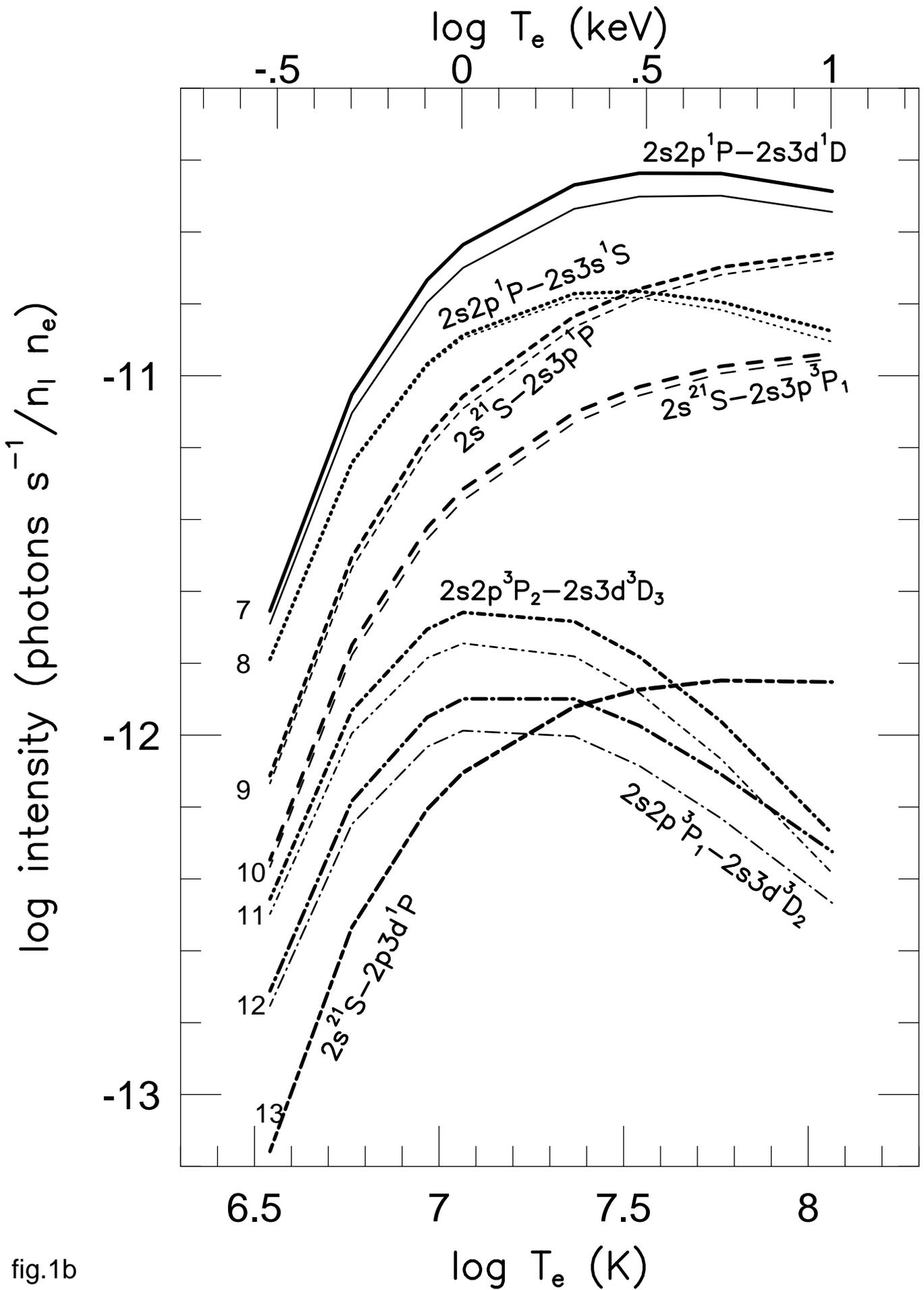
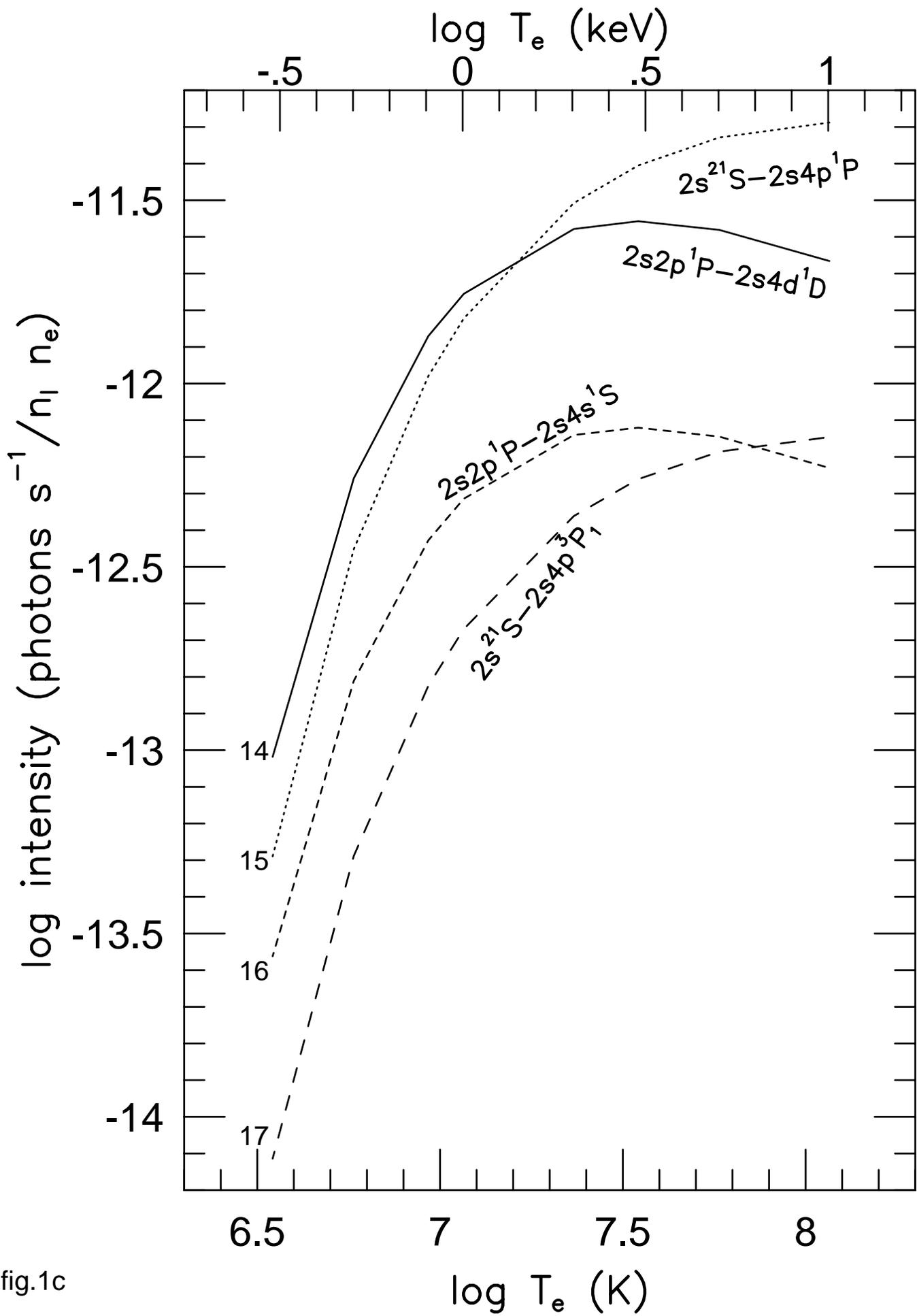
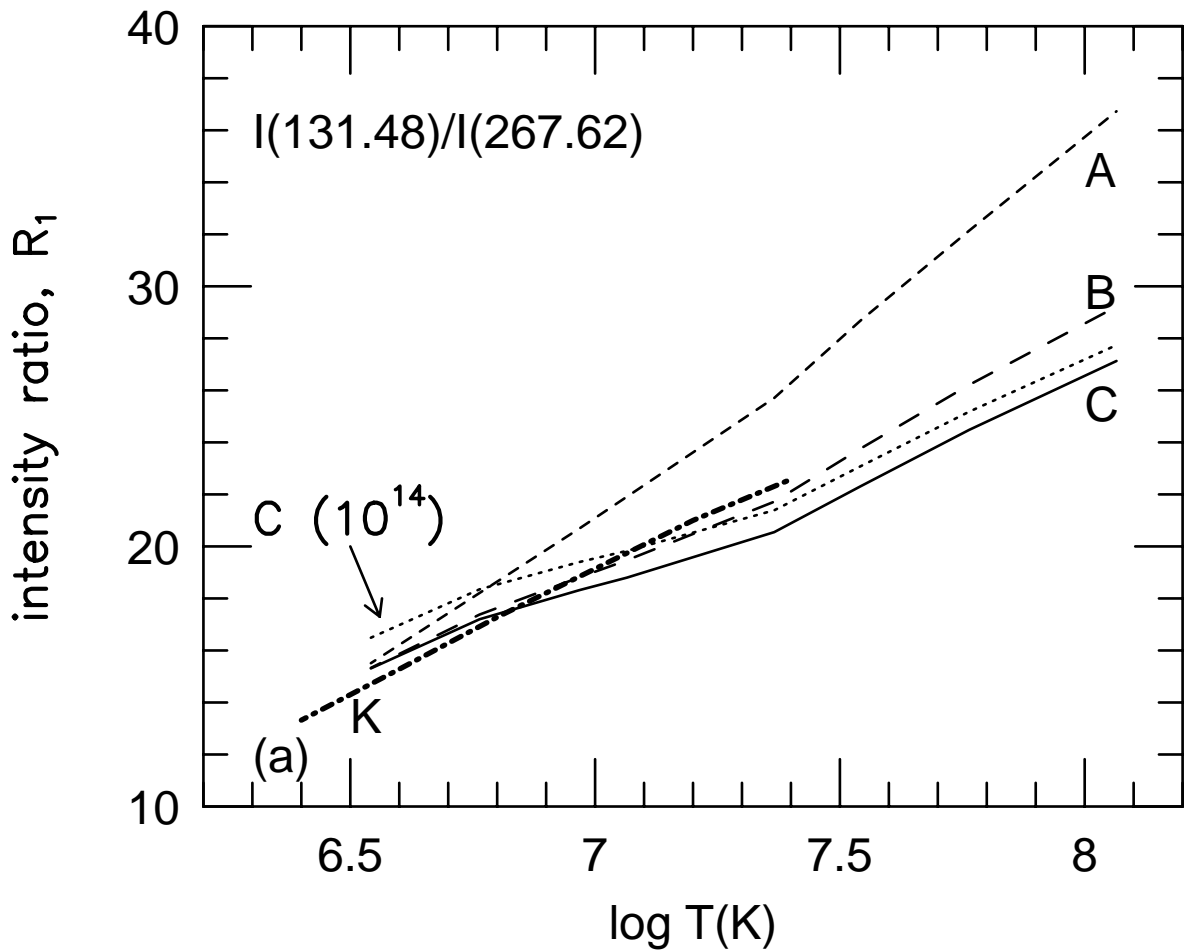
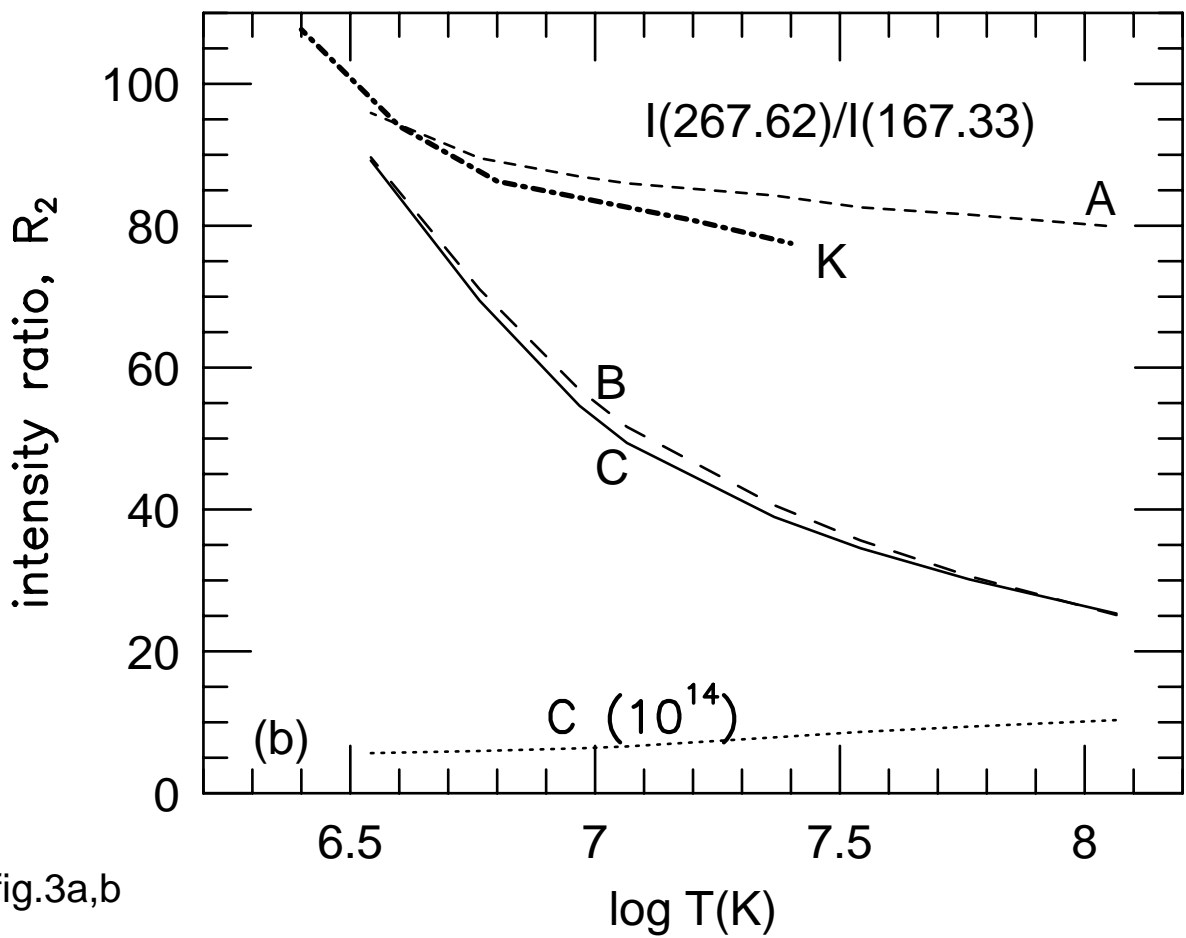
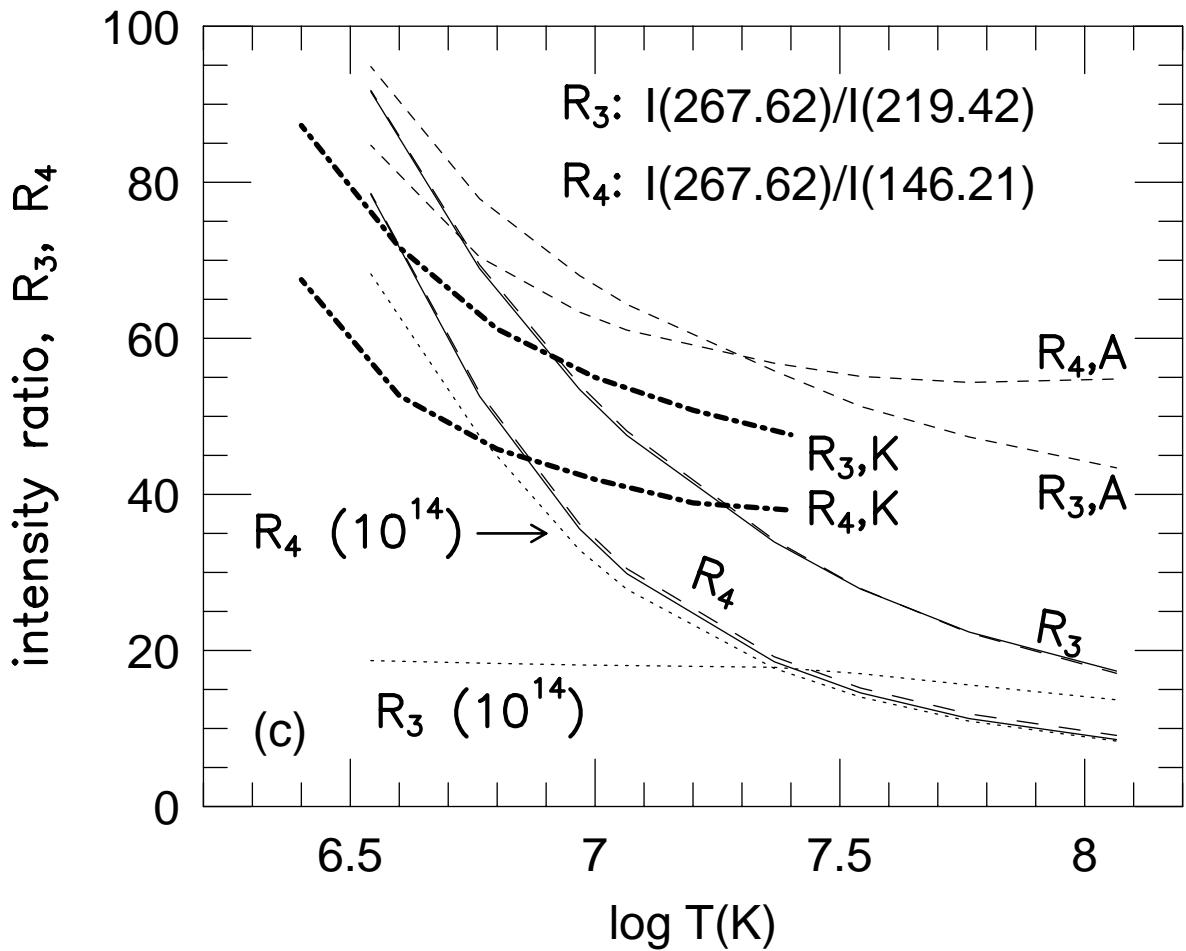
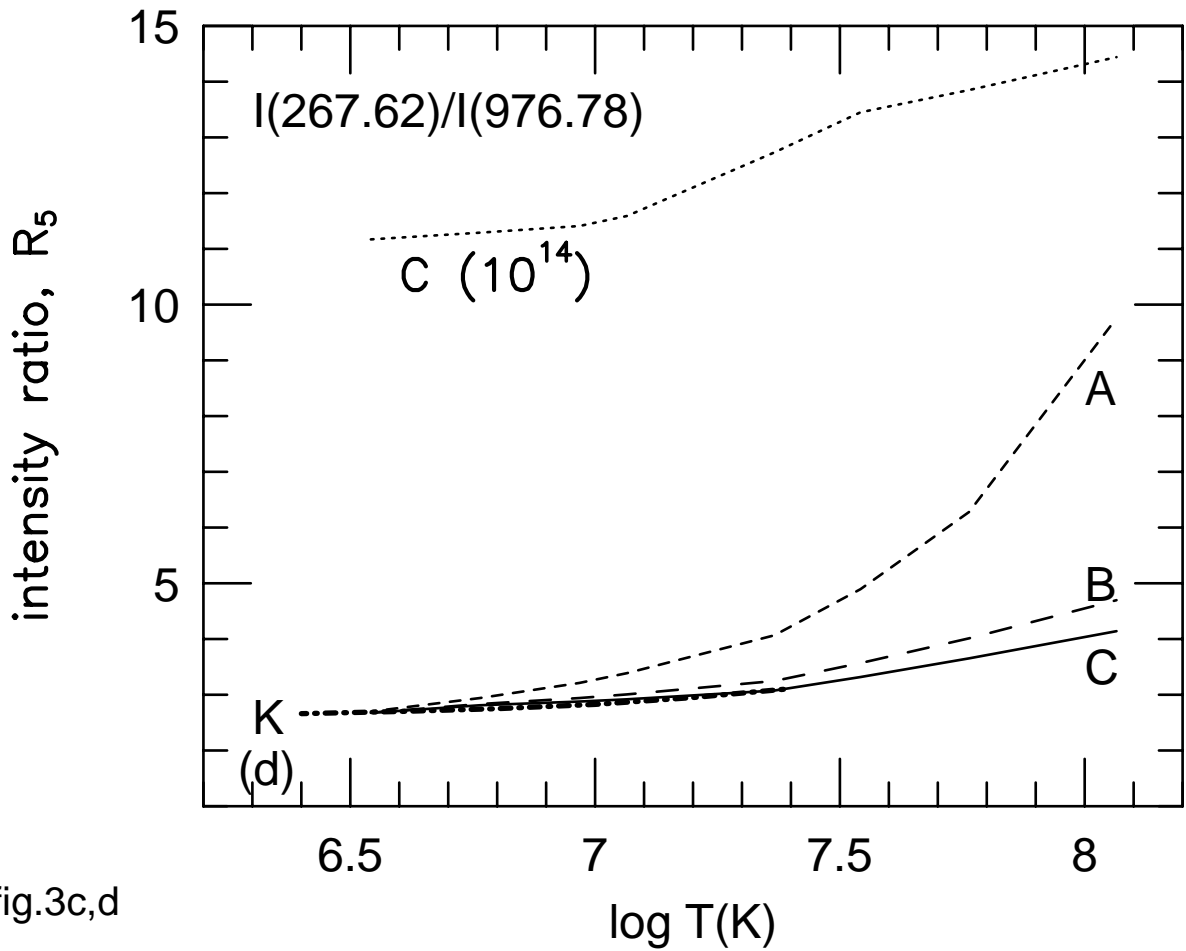
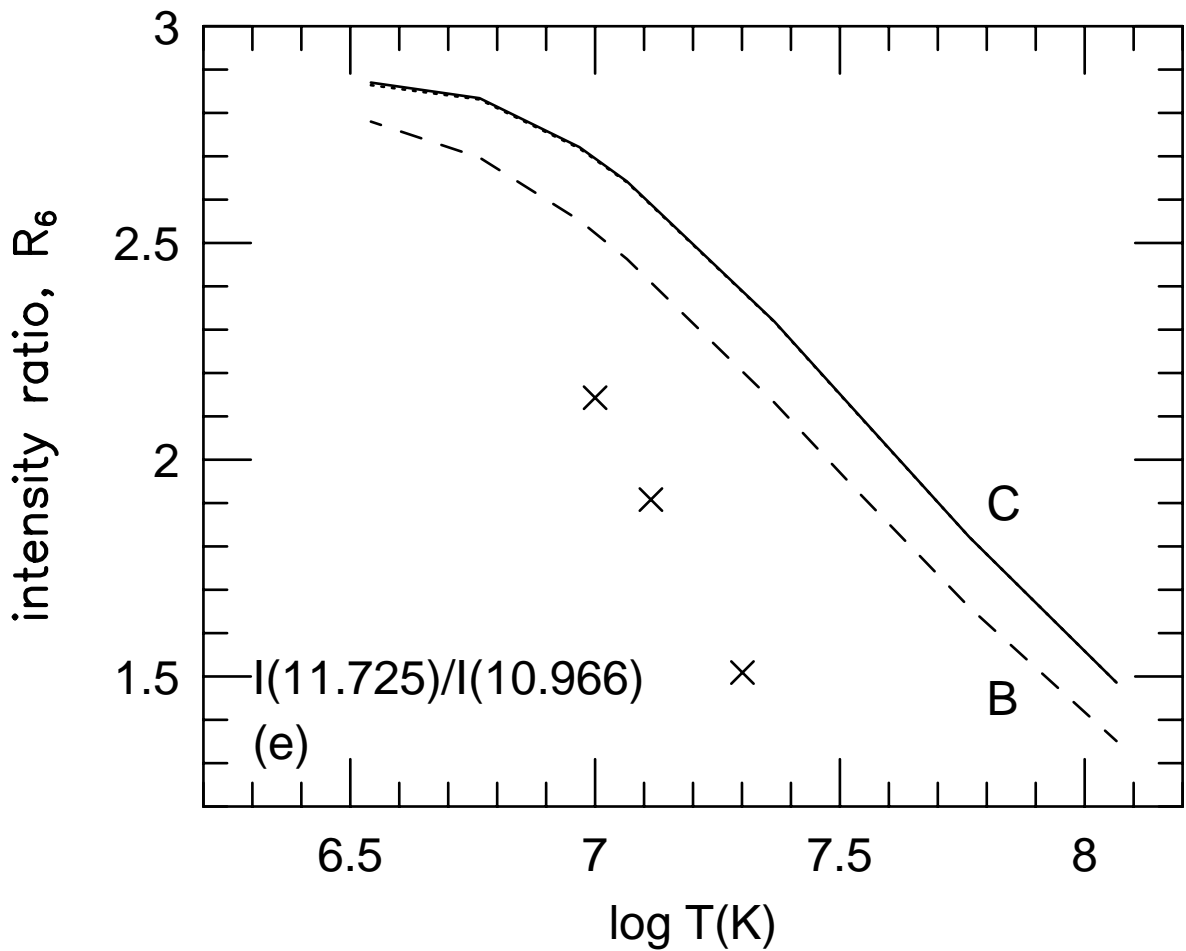
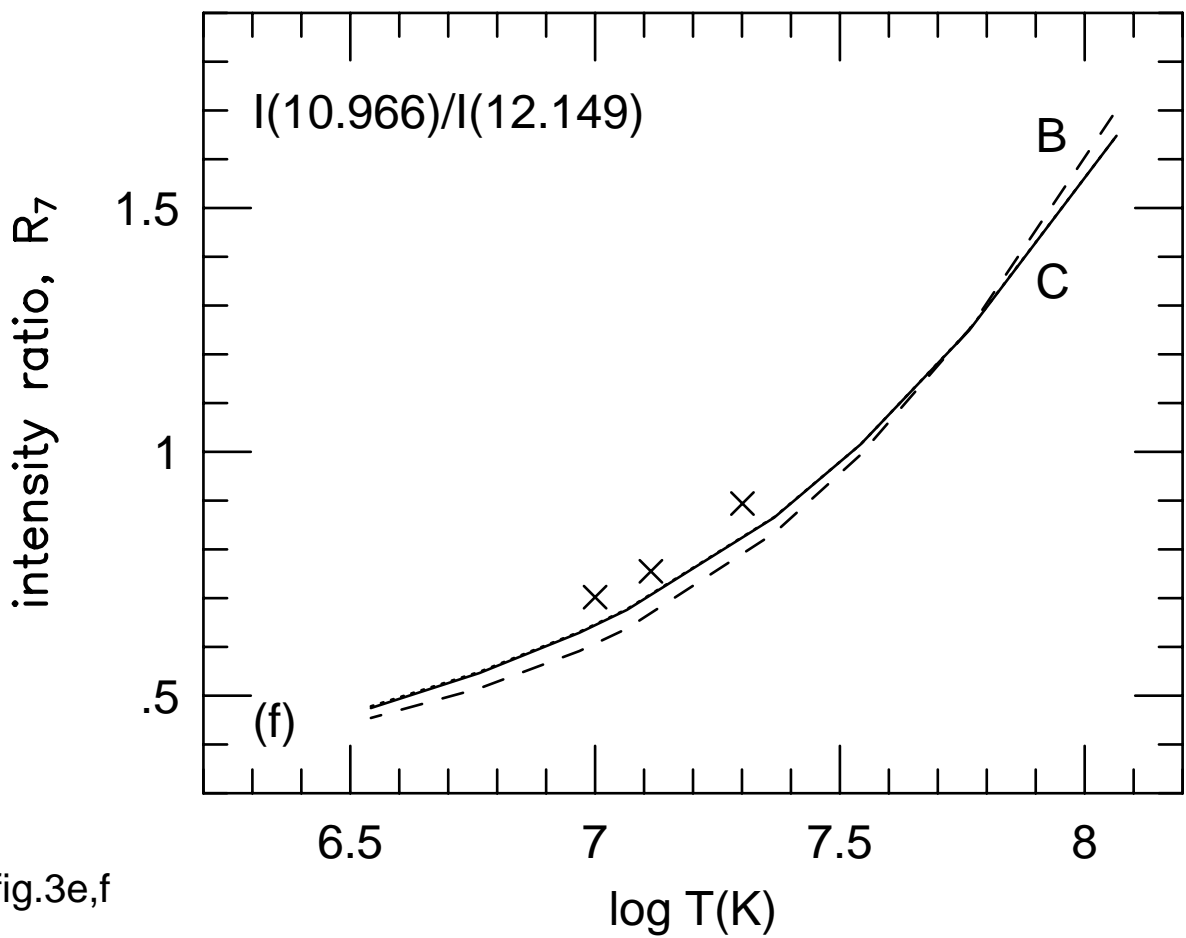
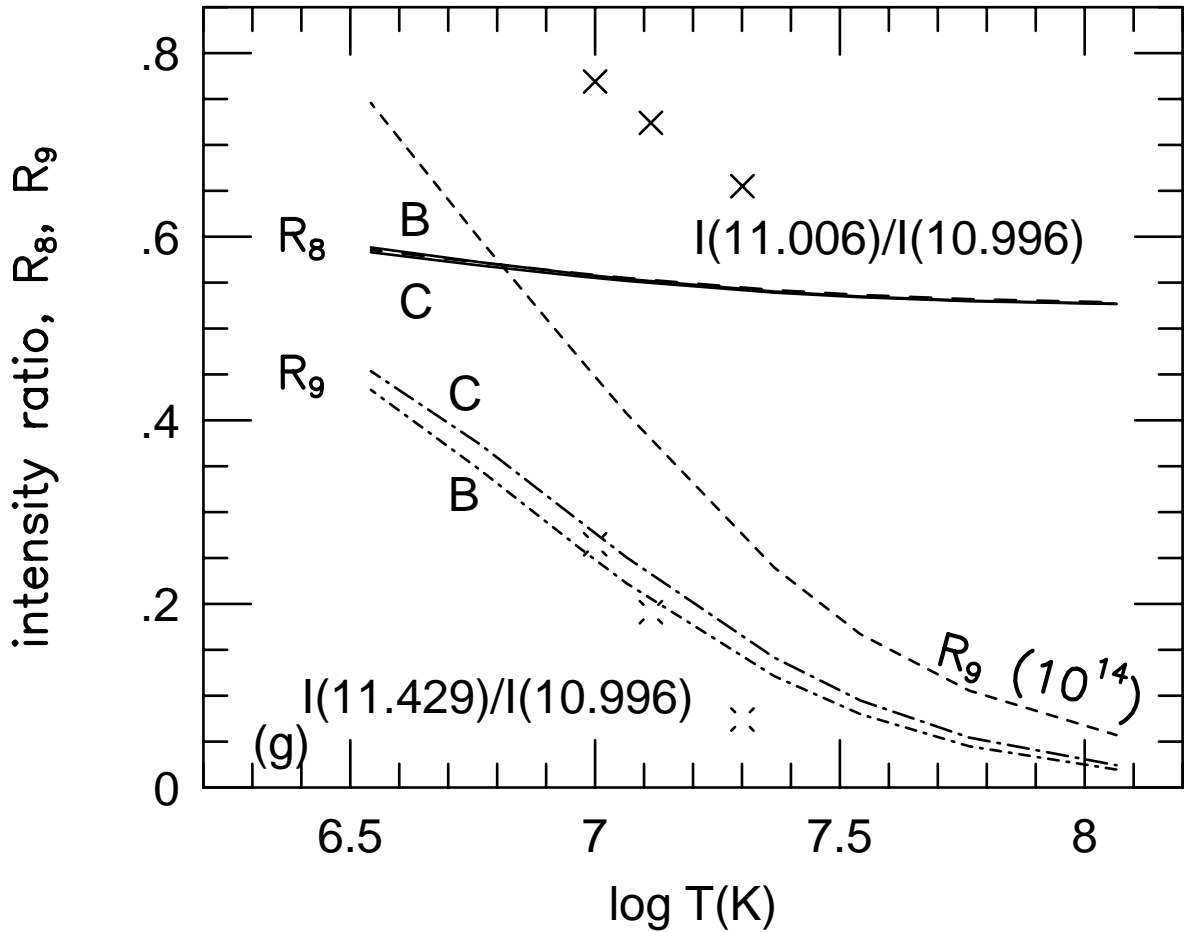
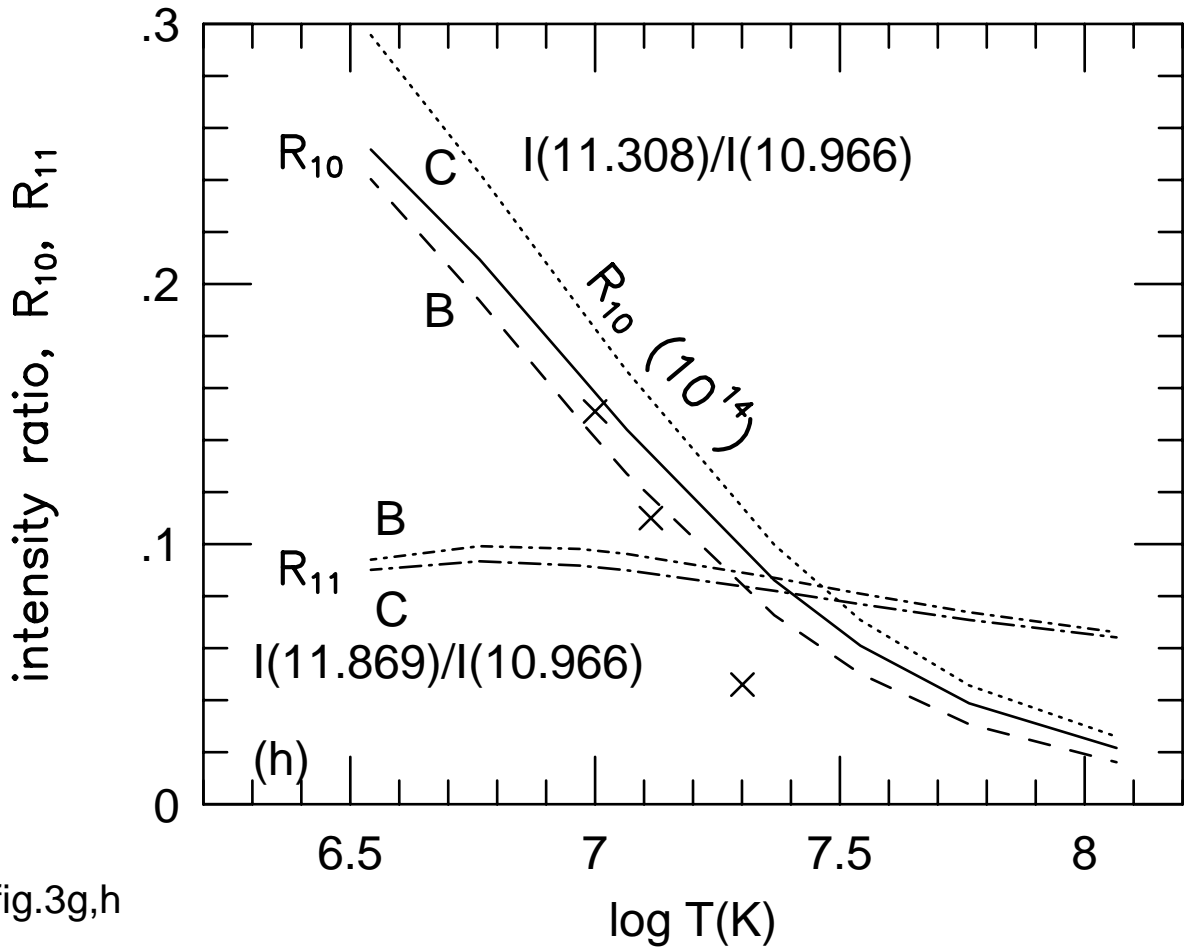
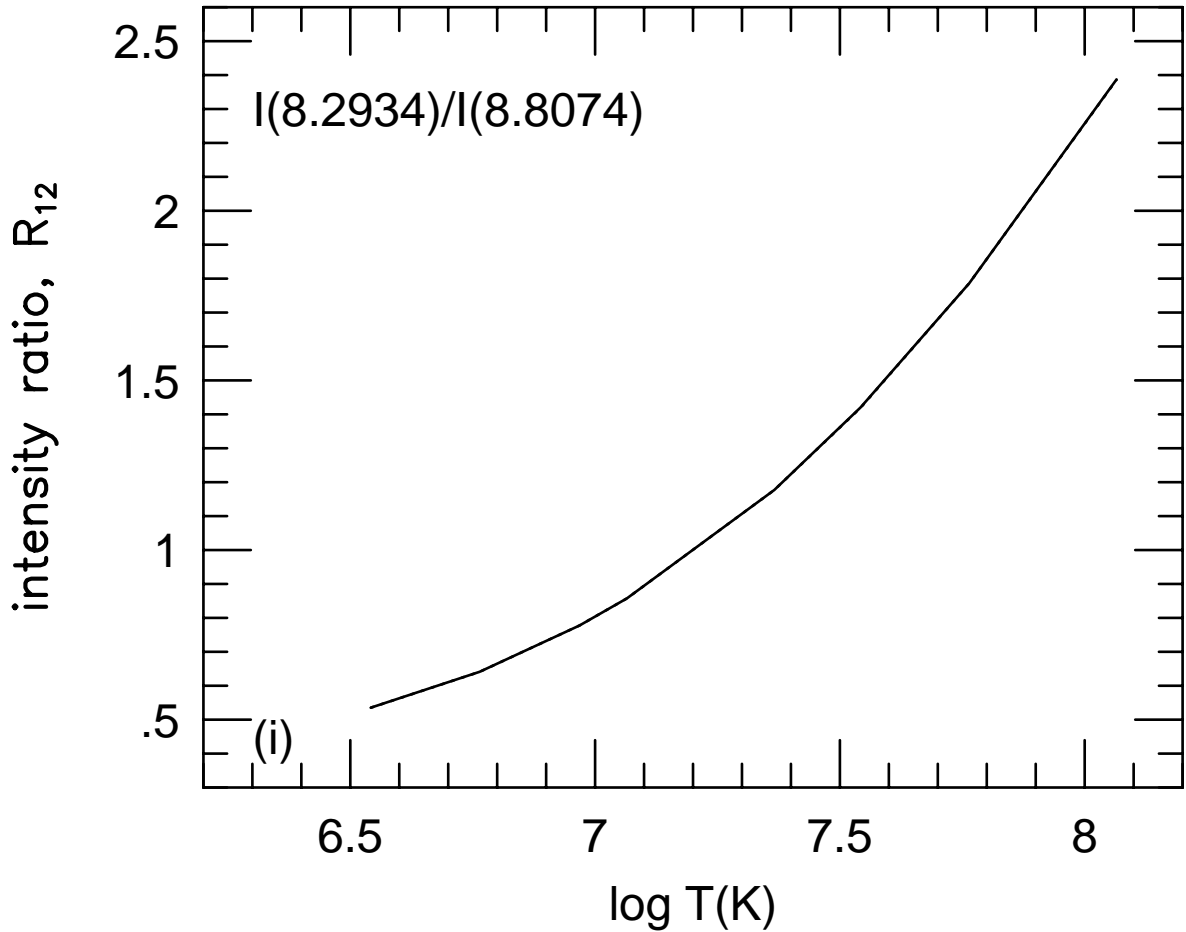
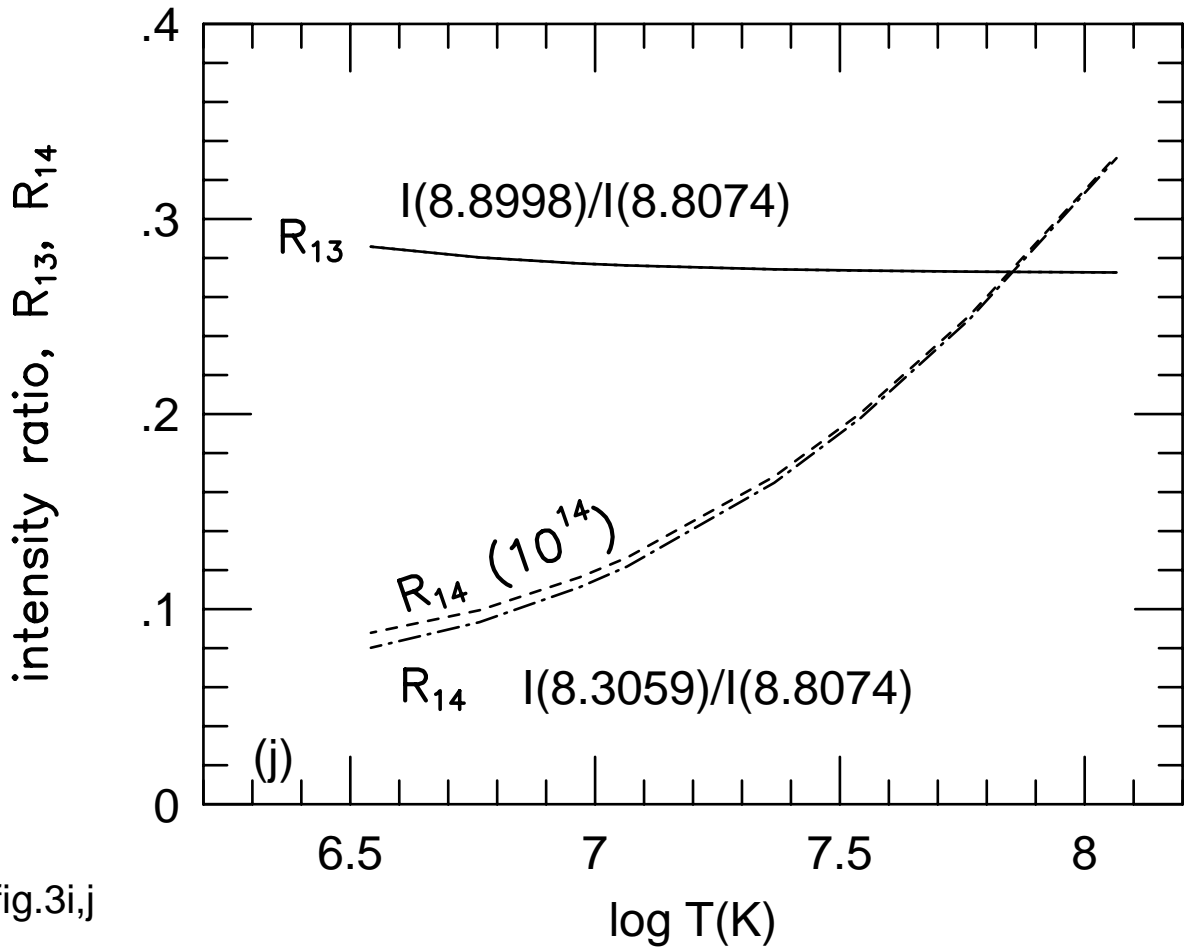
\end{document}